\documentclass[aps,preprint]{revtex4}
\usepackage{amsfonts}
\usepackage{amsmath}
\usepackage{amssymb}
\usepackage{subfigure}
\usepackage{graphicx}
\usepackage{epstopdf}
\usepackage{epsfig}
\usepackage{float}
\usepackage{hyperref}
\usepackage{color}
\usepackage{ulem}
\usepackage{cancel}

\begin{document}

\title{Phase Structures and Transitions of Quintessence Surrounding RN Black Holes in a Grand Canonical Ensemble}
\author{Yuchen Huang}
\email{huangyuchen@stu.scu.edu.cn}
\author{Hongmei Jing}
\email{hmjing@stu.scu.edu.cn}
\author{Jun Tao}
\email{taojun@scu.edu.cn}
\author{Feiyu Yao}
\email{yaofeiyu@stu.scu.edu.cn}

\affiliation{Center for Theoretical Physics, College of Physics, Sichuan University, Chengdu, 610065, China}

\begin{abstract}

Considering a grand canonical ensemble, we study the phase structures and transitions of RN black holes surrounded by quintessence dark energy on two different boundary conditions, namely AdS space and a Dirichlet wall. For AdS space, under the condition of fixed temperature and potential, as the temperature increases for lower potential, the black hole undergoes a first-order phase transition, while for higher potential, no phase transition occurs. There are two different regions in the parameter space. For the Dirichlet wall, on which the temperature and potential are fixed and the state parameter of quintessence $\omega=-2/3$ is analyzed in detail. Then, three different physically allowed regions in the parameter space of the black hole are well studied. As the temperature rises, a first-order phase transition and a second-order phase transition may occur. In this case, there are nine regions in the parameter space, which is obviously distinct from the case of AdS space.
\end{abstract}

\maketitle

\section{Introduction}

With the development of theory, people found a fundamental relationship between gravity, thermodynamics and quantum theory. Initially, physicists studied the mechanics of black holes \cite{Bardeen:1973gs}, and then they found that if they applied quantum field theory in curved spacetime to the fundamental properties of black holes, they could get the profound results. The framework of black hole thermodynamics is based on the discovery made by Bekenstein and Hawking that black holes have entropy and other thermodynamic quantities \cite{Bekenstein:1972tm,Bekenstein:1973ur,Hawking:1974sw,Bekenstein:1974ax}. The Schwarzschild black hole in AdS space was investigated by Hawking and Page \cite{Hawking:1982dh}, where the phase transition from thermal AdS to black hole was found. The black holes have negative specific heat in asymptotically flat spacetime which causes it to be an unstable system. The problem in asymptotically flat space then can be avoided since the gravitational potential of AdS space acts as a box of finite volume with unphysical perfectly reflecting walls \cite{Hawking:1982dh}. In addition, the emergence of AdS/CFT correspondence \cite{Gubser:1998bc,Witten:1998qj,Maldacena:1997re} has also inspired relevant researches on black hole thermodynamics \cite{Witten:1998zw,Chamblin:1999tk,Chamblin:1999hg,Caldarelli:1999xj,Cai:2001dz,Chen:2016gzz,Kubiznak:2012wp,Chen:2019nsr}.

Another preferred solution to solve this problem is to surround the black hole with a cavity. The Dirichlet boundary condition is supposed to be considered in the wall of the cavity.  This work was first carried out by York \cite{York:1986it}, in which the Hawking-Page-like phase transition was found. Subsequently, it was extended to the case of a RN black hole which was considered in a grand canonical ensemble \cite{Braden:1990hw} and a canonical ensemble \cite{Carlip:2003ne,Lundgren:2006kt}. The thermodynamic phases of diverse black holes in a cavity were discussed in \cite{Lu:2010xt,Wu:2011yu,Lu:2012rm,Lu:2013nt,Zhou:2015yxa,Xiao:2015bha,Wang:2019kxp,Wang:2019urm,Liang:2019dni,Wang:2020hjw}, which indicated that the Van de Waals-like phase transitions or the Hawking-Page-like phase transitions exist invariably except for some particular cases.
The relationship between thermodynamic properties of black holes and their specific boundary conditions in different extended phase spaces was studied in \cite{Zhao:2020nrx}. Apart from this, some other systems like charged scalars, boson stars and hairy black holes in a cavity were investigated in \cite{Basu:2016srp,Peng:2017gss,Peng:2017squ,Peng:2018abh}, which concluded that there are some surprising similarities between gravity systems in a cavity and holographic superconductors in the AdS gravity. The stabilities of solitons, stars and black holes in a cavity were also discussed in \cite{Sanchis-Gual:2015lje,Dolan:2015dha,Ponglertsakul:2016wae,Sanchis-Gual:2016tcm, Ponglertsakul:2016anb,Sanchis-Gual:2016ros,Dias:2018zjg,Dias:2018yey}. Recently, McGough, Mezei and Verlinde proposed that in the holographic dual, the $T\bar{T}$ deformed CFT$_{2}$ locates on a Dirichlet wall at finite radial distance of AdS$_3$ in the bulk \cite{McGough:2016lol}, which also inspires us to study the properties of black holes in a cavity.

From observations,  we know that universe is dominated by an energy component with an effective negative pressure \cite{Ostriker:1995su,Wang:1999fa}. One hypothesis for such a component is the cosmological constant, and other hypotheses are dynamical vacuum energy or quintessence \cite{Ratra:1987rm,Frieman:1991tu,Chiba:1997ej,Turner:1998ex,Shaisultanov:1997bc,Bucher:1998mh}. Quintessence is described by an ordinary scalar field minimally coupled to gravity, with particular potentials that lead to late time inflation. The quintessence field must couple to ordinary matter, and will lead to long range forces and time dependence of the constants of nature when it is suppressed by the Planck scale. The static spherically symmetric solution with the quintessential matter surrounding a charged black hole was first presented in \cite{Kiselev:2002dx}. Since then, a series of works about black holes in presence of quintessence had been done. Quintessence modes were studied in \cite{Chen:2005qh,Zhang:2006ij}, thermodynamics including Hawking radiation, heat engine, phase transition in the holographic framework and so on were investigated in \cite{Chen:2008ra, Wei:2011za, AzregAinou:2012hy, Li:2014ixn, Zeng:2015wtt, Ma:2016arz, Liu:2017baz,He:2019fti,Hendi:2020ebh,Yan:2021}. Especially, the phase transition of the quintessence RN-dS black hole was studied through the effective thermodynamic quantities \cite{Liu:2019qxt}. Although many relevant work has been completed, there is no research that combines quintessential matter and cavity boundary condition. In addition, the discovery about the difference of phase structures of black holes between AdS space and Dirichlet cavity \cite{Wang:2019kxp,Wang:2019urm,Liang:2019dni,Wang:2020hjw} also prompts us to investigate the phase structures of black holes from another perspective.

In this paper, we study the phase structures and transitions of RN black holes surrounded by quintessence dark energy under two different boundary conditions, i.e., AdS space and a Dirichlet wall. Besides, a grand canonical ensemble is considered, which indicates that the temperature and potential of the black hole are fixed on AdS boundary condition while the temperature and potential of the cavity wall are fixed on Dirichlet wall boundary condition. The paper is organized as follows: In section II, we discuss the phase structures and transitions of a RN-AdS black hole surrounded by quintessence dark energy. In section III, we study the phase structures and transitions of RN black holes surrounded by quintessence dark energy in a cavity. The final summaries and discussions are presented in section IV.

\section{Quintessence RN Black Holes In AdS Space}

In this section, we discuss the phase structures and transitions of the RN black hole surrounded by quintessence dark energy in AdS space. The action of the black hole in four dimensional curved space-time is \cite{Ghaffarnejad:2018bsd}
\begin{equation}
\mathcal{S}=\frac{1}{16\pi}\int d^4 x(\sqrt{-g}[R-2\Lambda-F^{\mu\nu}F_{\mu\nu}]+ \mathcal{L}_{q}). \label{action}
\end{equation}
In the above action, the cosmological constant $\Lambda=-3/l^2$, where $l$ is the AdS space radius. And $\mathcal{L}_{q}$ is Lagrangian of quintessence as a barotropic perfect fluid, which can be written as \cite{Minazzoli:2012md}
\begin{equation}
\mathcal{L}_{q}=-\rho_q\left[1+\omega ln(\frac{\rho_{q}}{\rho_0})\right], \label{Lq}
\end{equation}
where $\rho_{q}$ is the energy density, $\rho_{0}$ is an integral constant, and $\omega$ is the barotropic index.
The solution of the spherical symmetric charged quintessence RN-AdS black hole is given by
\begin{equation}
ds^2=f(r)dt^2-\frac{dr^2}{f(r)}-r^2(d\theta^2+sin^2 \theta d\phi^2),
\end{equation}
and
\begin{equation}
f(r)=1-\frac{2M}{r}+\frac{Q^2}{r^2}-\frac{a}{r^{3\omega+1}}+\frac{r^2}{l^2},
\end{equation}
where $M$ and $Q$ are the mass and charge of the black hole, respectively.
The normalization factor $a$ is related to the density of quintessence dark energy
\begin{equation}
\rho_{q}=-\frac{a}{2}\frac{3\omega}{r^3(\omega+1)},
\end{equation}
here $a$ needs to be positive and the constraint on the barotropic index is $-1<\omega<-1/3$.

Since the horizon raidus $r_h$ is the root of $f(r_h)=0$, we have
\begin{equation}
M=\frac{r_{h}}{2}+\frac{Q^2}{2r_{h}}-\frac{a}{2r_{h}^{3\omega}}+\frac{r_{h}^3}{2l^2}. \label{M}
\end{equation}
In the discussion that follows, $M$ is redundant so we replace it with $r_{h}$. Thus the Hawking temperature is given by
\begin{equation}
T=\frac{f^{'}(r_{h})}{4\pi}=\frac{1}{4\pi r_{h}}\left(1+\frac{3r_{h}^2}{l^2}-\frac{Q^2}{r_{h}^2}+\frac{3a\omega}{r_{h}^{3\omega+1}}\right).\label{T}
\end{equation}

The potential and entrophy of the black hole are
\begin{equation}
\Phi=\frac{Q}{r_h}, \text{ } S=\pi r_h^2. \label{Phi}
\end{equation}
In addition, free energy plays a crucial role in probing into the phase structures and transitions of quintessence RN-AdS black hole in grand canonical ensemble and it's given by \cite{Liang:2019dni,Wang:2018xdz}
\begin{equation}
F=M-TS-Q\Phi, \label{F}
\end{equation}
in which the mass of the black hole is interpreted as enthalpy \cite{Kubiznak:2016qmn}. Ignoring the variation of $l$, we express physical quantities in units of $l$ from now on to simplify the later calculation
\begin{equation}
	\widetilde{r}_{h}=r_h/l,\ \widetilde{a}=a/l^{3\omega+1},\ \widetilde{Q}=Q/l,\ \widetilde{T}=Tl,\ \widetilde{F}=F/l.
\end{equation}

\begin{figure}[ptb]
\begin{center}
\includegraphics[width=0.5\textwidth]{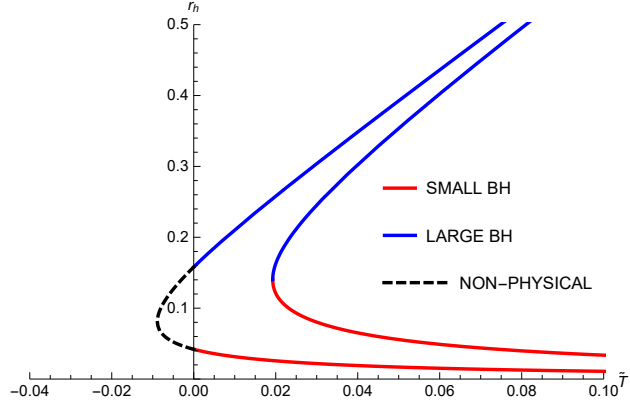}
\caption{~\scriptsize Plot of $\widetilde{r}_h$ against $\widetilde{T}$ for different $\Phi$ with $\omega=-2/3$ and $\widetilde{a}=0.3$, where $\Phi=0.99, 0.97$ from left to right. }
\label{1r-T}
\end{center}
\end{figure}

The heat capacity at constant potential is
\begin{equation}
C_{\Phi}=T\left(\frac{\partial S}{\partial T}\right)_{\Phi}=2\pi l^2 \widetilde{r}_{h} \widetilde{T} (\frac{\partial \widetilde{r}_{h}}{\partial \widetilde{T}})_{\Phi}. 
\end{equation}
Thus the sign of the heat capacity is the same as the sign of $\partial \widetilde{r}_h/\partial \widetilde{T}$ at constant potential. We eliminate $Q$ by Eq. (\ref{T}) and Eq. (\ref{Phi}) to obtain $\widetilde{T} = \widetilde{T}\left(\widetilde{r}_{+}, \Phi \right)$
\begin{equation}
	\widetilde{T}=\frac{1}{4\pi \widetilde{r}_{h}}\left(1-\Phi^2+3\widetilde{r}_{h}^2+\frac{3\widetilde{a}\omega}{\widetilde{r}_{h}^{3\omega+1}}\right).\label{T2}
\end{equation}
According to Eq. (\ref{T2}), curves of the horizon radius $\widetilde{r}_{h}$ in terms of temperature $\widetilde{T}$ with fixed values of $\Phi$ are plotted in FIG. \ref{1r-T}. For a fixed the potential $\Phi$, $\widetilde{T}\left(\widetilde{r}_{h}, \Phi \right)$ has a minimum $\widetilde{T}_{min}$. When $\widetilde{T}>\widetilde{T}_{min}$, there are two branches, among which the Small BH with $\partial\widetilde{r}_h/\partial\widetilde{T}<0$ is thermodynamically unstable and the Large BH with $\partial\widetilde{r}_h/\partial\widetilde{T}>0$ is thermodynamically stable. There exists one section of non-physical curve in FIG. \ref{1r-T} since temperature can't be negative.

\begin{figure}
\begin{center}
\includegraphics[width=0.6\textwidth]{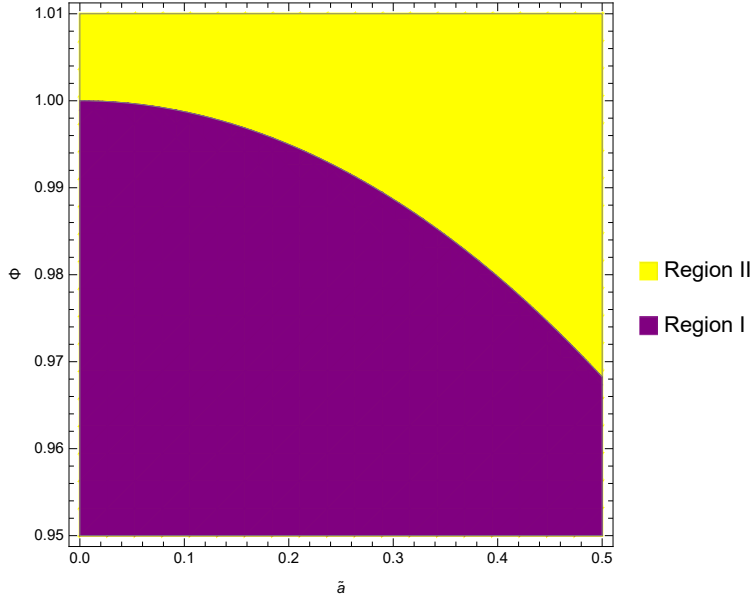}
\caption{~\scriptsize Two regions in $\Phi-\widetilde{a}$ phase space with $\omega=-2/3$.} \label{1phi-a}
\end{center}
\end{figure}

It's worth noting that the thermal AdS space with a constant gauge potential is a classical solution of action (\ref{action}) in the grand canonical ensemble, which should be also considered for our discussion. We find that there are two regions in $\Phi-\widetilde{a}$ phase space. As shown in FIG. \ref{1phi-a}, these two regions are referred to Region I and Region II. Specifically, in Region I, there exists a first-order phase transition as temperature $\widetilde{T}$ increases. Nevertheless, in Region II, there is no phase transition.

\begin{figure}[ptb]
\begin{center}
\subfigure[{~\scriptsize Region I: $\omega=-2/3$, $\widetilde{a}=0.3$, $\Phi=0.97$.}]{
\includegraphics[width=0.45\textwidth]{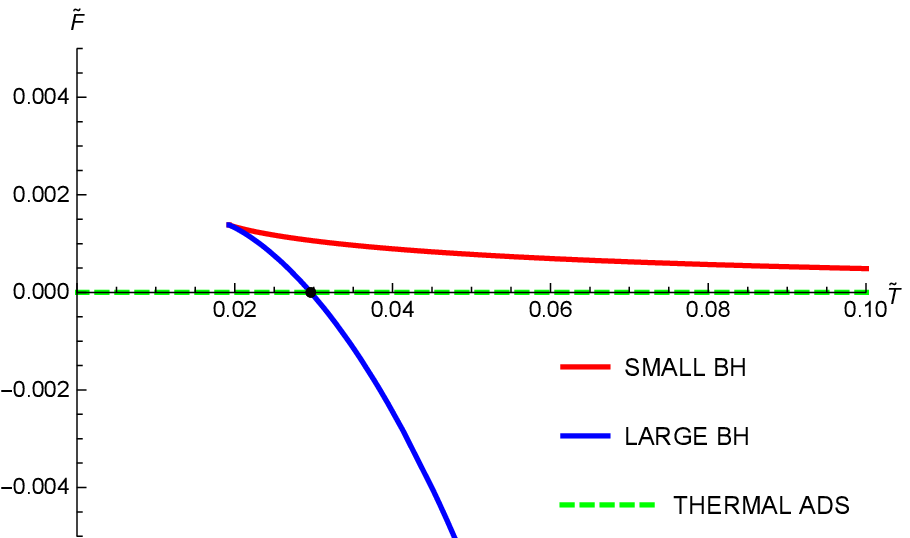}\label{1F-T1}}
\subfigure[{~\scriptsize Region II: $\omega=-2/3$, $\widetilde{a}=0.3$, $\Phi=0.99$.}]{\includegraphics[width=0.45\textwidth]{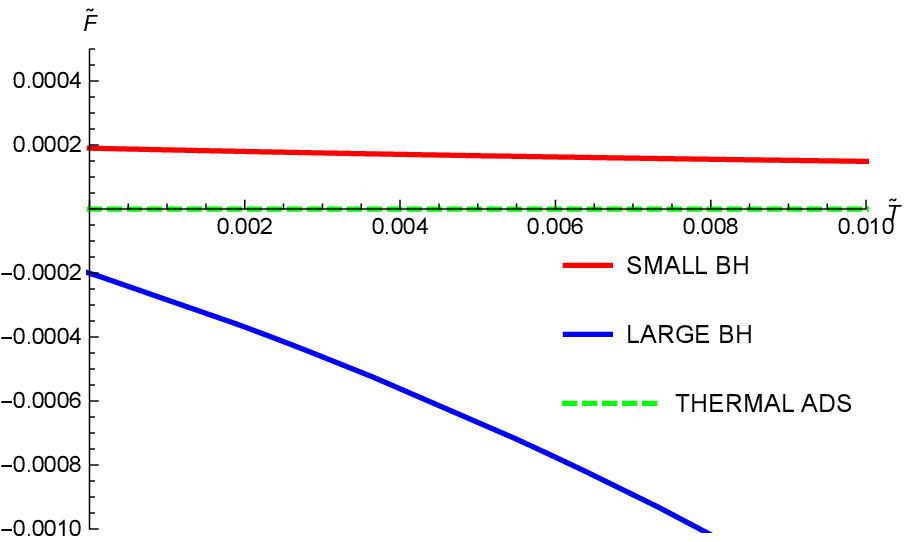}\label{1F-T2}}
\caption{~\scriptsize Plots of free energy $\widetilde{F}$ against temperature $\widetilde{T}$ for fixed $\Phi$ of two regions. The black dot is where a first-order phase transition occurs.}
\label{1F-T}
\end{center}
\end{figure}

We discuss the phase structures and transitions of the black hole in the abovementioned two regions. Our method is as follows. Eq. (\ref{M}) and Eq. (\ref{T}) imply that $M$ and $T$ are functions of $r_{h}$ and $Q$. According to Eq. (\ref{Phi}), the charge can be written as the function of $r_{h}$ and $\Phi$, i.e. $Q=Q(r_{h},\Phi)$. Substituting it into $M(r_{h},Q)$ and $T(r_{h},Q)$, we have $M(r_{h},\Phi)$ and $T(r_{h},\Phi)$. Thus $F$ can be written as the function of $r_{h}$ and $\Phi$ according to Eq. (\ref{F}). Quantity in units of $l$ can be written as $\widetilde{F}(\widetilde{r}_{h},\Phi)$. The $\widetilde{F}-\widetilde{T}$ diagrams with fixed $\Phi$ are shown in FIG. \ref{1F-T} by regarding $\widetilde{r}_{h}$ as the parameter. We notice that the behavior of the black hole for each value of $\Phi$ is self-consistent with that shown in FIG. \ref{1r-T}, that is, for $\widetilde{T}>\widetilde{T}_{min}$, there are always two branches, i.e. Small BH and Large BH, it shows again that the Large BH is more stable than the Small BH because of the lower free energy. In Region I, when temperature $\widetilde{T}$ increases from zero but doesn't exceed $\widetilde{T}_{min}$, the Thermal AdS is the only phase. When $\widetilde{T}>\widetilde{T}_{min}$, the Small BH branch and Large BH branch appear but the Thermal AdS is still the globally stable state until the occurrence of a first-order phase transition. As $\widetilde{T}$ increases further, the Large BH becomes the thermodynamically preferred state ultimately. This process is illustrated in FIG. \ref{1F-T1}. In Region II, there is no phase transition. Apparently, for any given temperature, Large BH is always the globally stable phase, just as shown in FIG. \ref{1F-T2}.

\begin{figure}[ptb]
\begin{center}
\subfigure[{~\scriptsize $\widetilde{a}=0.3$.}]{
\includegraphics[width=0.4\textwidth]{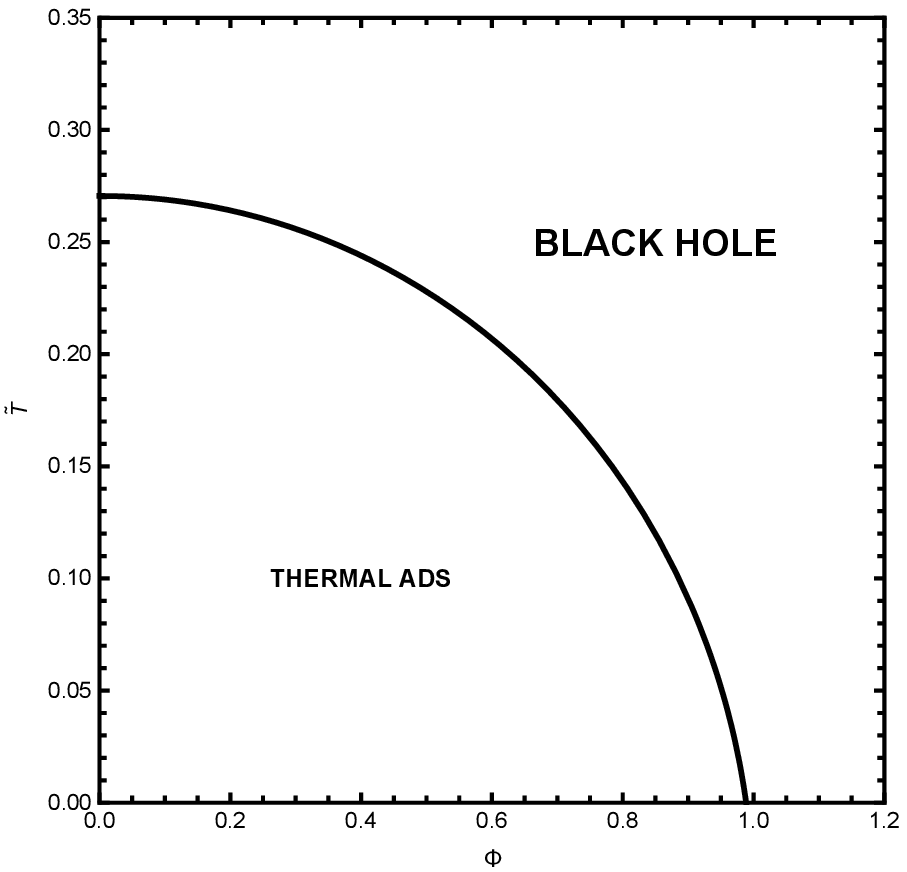}\label{1T-phi1}}
\subfigure[{~\scriptsize $\widetilde{a}=1$.}]{
\includegraphics[width=0.4\textwidth]{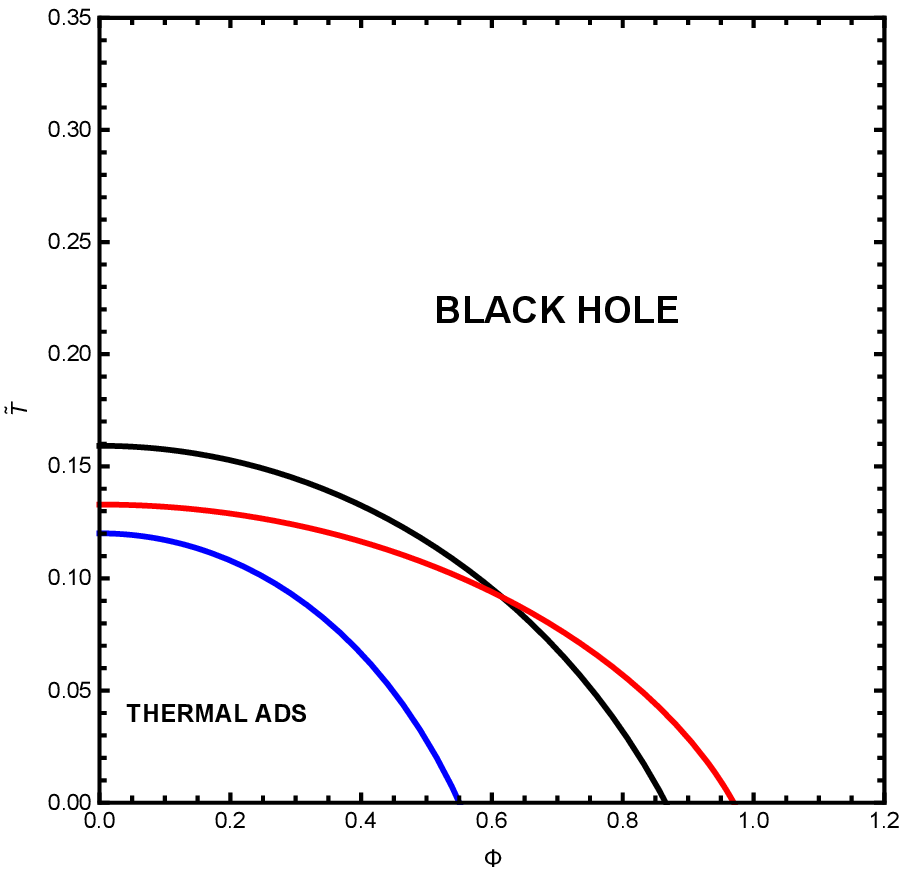}\label{1T-phi2}}
\caption{~\scriptsize The phase diagrams in $\widetilde{T}-\Phi$ space for various $\omega$ and $\widetilde{a}$. Blue curve: $\omega=-2/5$. Black curve: $\omega=-2/3$. Red curve: $\omega=-9/10$. The first-order phase transitions occur on these curves, causing them to become the dividing lines of two phases.}
\label{1T-phi}
\end{center}
\end{figure}

Another $\widetilde{T}-\Phi$ phase diagram is displayed in FIG. \ref{1T-phi}. To avoid misunderstandings, the Small BH and Large BH are both denoted as Black Hole. We keep the parameters fixed, that is, set $\omega=-2/3$ and $\widetilde{a}=0.3$ in FIG. \ref{1T-phi1}. The temperature of the phase transition decreases to zero as $\Phi$ rises. Another homologous phase diagram is shown in FIG. \ref{1T-phi2}, in which we adjust $\omega = -2/5, -2/3, -9/10$ respectively and $\widetilde{a} = 1$. Compared with the former, this case is slightly distinct but the similarity is that there is only a first-order phase transition from the Thermal AdS to Black Hole.

\section{Quintessence RN Black Holes In A Cavity}

In this section, we investigate the RN black hole surrounded by quintessence dark energy in a cavity. The action with a time-like boundary $\partial \mathcal{M}$ on a four dimensional spacetime maniford $\mathcal{M}$ is
\begin{equation}
\mathcal{S}=\frac{1}{16\pi}\int_{\mathcal{M}}d^4 x\left(\sqrt{-g}[R-F^{\mu\nu}F_{\mu\nu}]+\mathcal{L}_{q}\right)-\frac{1}{8\pi}\int_{\partial \mathcal{M}}d^3 x\sqrt{-\gamma}\left(K-K_0\right). \label{action2}
\end{equation}
The counter term above is the Gibbons-Hawking-York surface term. $K$ is the
extrinsic curvature, $\gamma$ is the metric on the boundary, and $K_{0}$ is a subtraction term to make the Gibbons-Hawking-York term vanish in flat spacetime. When the metric on $\partial \mathcal{M}$ is fixed, the Gibbons-Hawking-York term is crucial to obtain the correct the equations of motion from performing the variation \cite{Wang:2019kxp}. And we can get the quintessence RN black hole solution from the action (\ref{action2})
\begin{equation}
ds^2=f(r)dt^2-\frac{dr^2}{f(r)}-r^2(d\theta^2+sin^2 \theta d\phi^2),
\end{equation}
and
\begin{equation}
f(r)=1-\frac{2M}{r}+\frac{Q^2}{r^2}-\frac{a}{r^{3\omega+1}}. \label{metric2}
\end{equation}

Suppose that the wall of the cavity surrounding the black hole is located at $r_{B}$. For such a system in a grand canonical ensemble, the wall is maintained at a temperature $T$ and a potential $\Phi$. The charge $Q$ of the system is defined equal to the black hole's charge without cavity and the entrophy $S$ of the system is defined as equal to a quarter of the  event horizon area of the black hole $S=\pi r_{+}^2$. For such a system, the total energy can be written as \cite{Brown:1994gs}
\begin{equation}
	E=r_B\left[1-\sqrt{f(r_B)}\right].
\end{equation}
We define the temperature equal to
\begin{equation}
	T=\left(\frac{\partial E}{\partial S}\right)_{Q}=\frac{T_h}{\sqrt{f(r_B)}}=\frac{1}{4\pi r_+ \sqrt{f(r_B)}}\left(1-\frac{Q^2}{r_+^2}+\frac{3a\omega}{r_+^{3\omega+1}}\right),
\end{equation}
where $T_h$ is the Hawking temperature of the black hole and we have used $r_+$ to eliminate $M$
\begin{equation}
	M=\frac{r_{+}}{2}+\frac{Q^2}{2r_{+}}-\frac{a}{2r_{+}^{3\omega}}.\label{M}
\end{equation}
And the potential is defined to
\begin{equation}
	\Phi=\left(\frac{\partial E}{\partial Q}\right)_{S}=\frac{Q\left(\frac{1}{r_+}-\frac{1}{r_B}\right)}{\sqrt{f\left(r_B\right)}}.
\end{equation}
With all these quantities above, the first law of thermodynamics thus can be written as
\begin{equation}
	dE=TdS+\Phi dQ.
\end{equation}
The free energy can be given by
\begin{equation}
	F=E-TS-\Phi Q.\label{G}
\end{equation}
Physical quantities in units of $r_B$ can be concise and to the point
\begin{equation}
	x=r_+/r_B,\  \widetilde{a}=a/r_{B}^{3\omega+1},\  \widetilde{Q}=Q/r_B,\  \widetilde{T}=16\pi r_B T,\  \widetilde{F}=F/r_B.\label{pqurb}
\end{equation}
Plugging Eq. (\ref{pqurb}) into Eq. (\ref{G}), we can get the specific expression for free energy
\begin{equation}
	\widetilde{F}=1-\sqrt{1-\widetilde{a}-x+\widetilde{Q}^2-\frac{\widetilde{Q}^2}{x}+\frac{\widetilde{a}}{x^{3\omega}}}-\frac{\widetilde{T} x^2}{16}-\Phi \widetilde{Q}. \label{Gibbsfreeenergy}
\end{equation}
Similarly, temperature and potential can be written in units of $r_B$
\begin{equation}
\widetilde{T}=\frac{4}{ x\sqrt{1-\widetilde{a}-x+\widetilde{Q}^2-\frac{\widetilde{Q}^2}{x}+\frac{\widetilde{a}}{x^{3\omega}}}}\left(1-\frac{\widetilde{Q}^2}{x^2}+\frac{3\widetilde{a}\omega}{x^{3\omega+1}}\right) \label{Tinunitsofrb}
\end{equation}
and
\begin{equation}
\Phi=\frac{\widetilde{Q}(1/x-1)}{\sqrt{1-\widetilde{a}-x+\widetilde{Q}^2-\frac{\widetilde{Q}^2}{x}+\frac{\widetilde{a}}{x^{3\omega}}}}. \label{Phiinunitsofrb}
\end{equation}

\begin{figure}[ptb]
\begin{center}
\subfigure[{~\scriptsize $0<\widetilde{a}\leq1/3$. H State and M State are candidates with the globally minimal free energy.}]{
\includegraphics[width=0.32\textwidth]{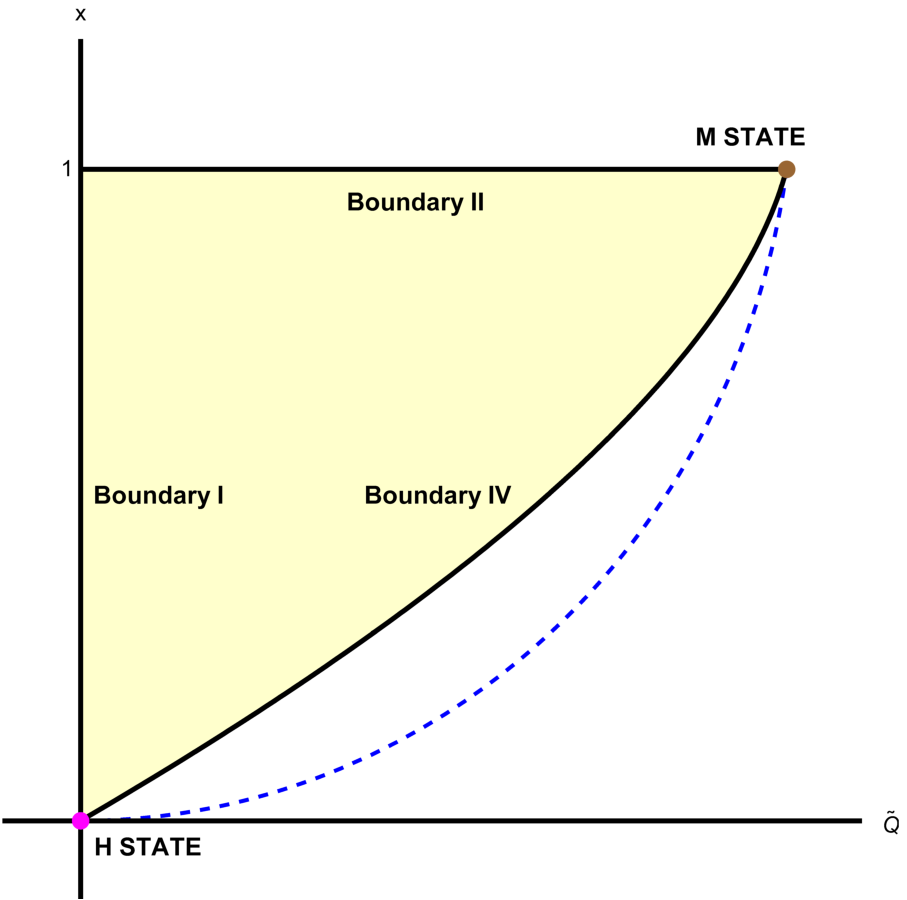}\label{2x-Q1}}
\subfigure[{~\scriptsize $1/3<\widetilde{a}<1/2$. H State and L State are candidates with the globally minimal free energy.}]{
\includegraphics[width=0.32\textwidth]{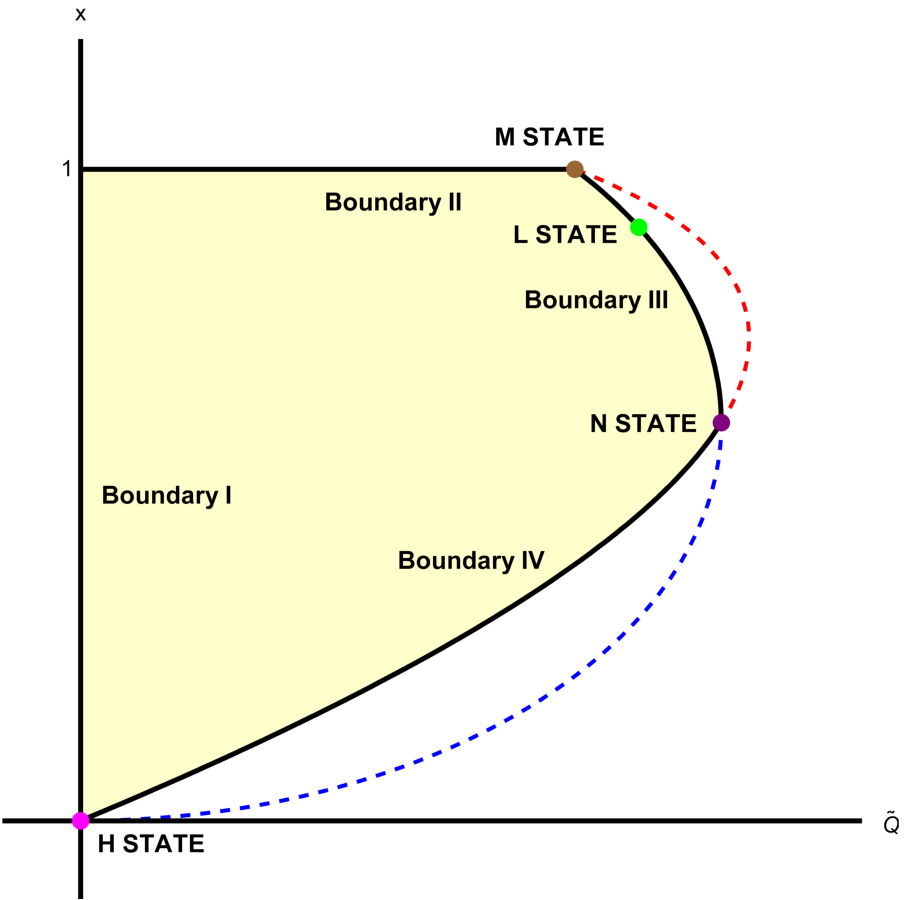}\label{2x-Q2}}
\subfigure[{~\scriptsize $1/2\leq\widetilde{a}<1$. H State and L State are candidates with the globally minimal free energy.}]{
\includegraphics[width=0.32\textwidth]{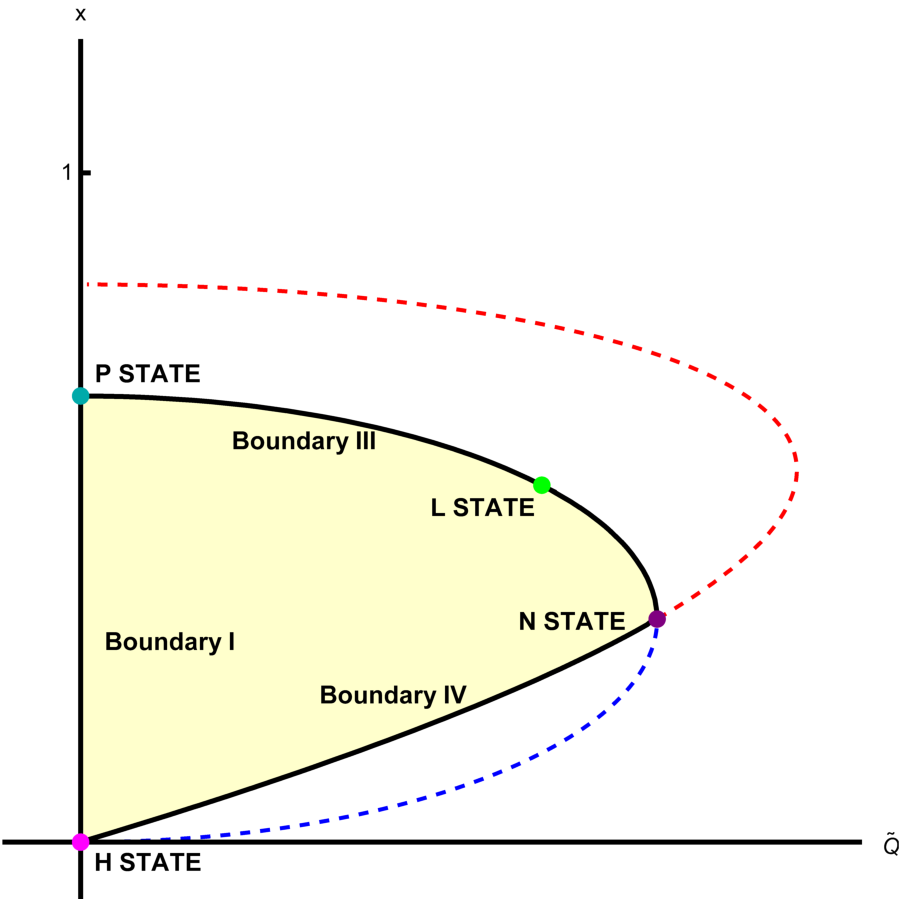}\label{2x-Q3}}
\caption{~\scriptsize Three types of distinct physically allowed regions in $x-\widetilde{Q}$ space. The yellow area is physically allowed and the black curve is the boundary. The coordinates of these states are: H State(0,0), M State($\sqrt{1-2\widetilde{a}}$,1), N State($\frac{1}{2} \sqrt{\widetilde{a}+\frac{1}{\widetilde{a}}-2},\frac{1-\widetilde{a}}{2 \widetilde{a}}$), L State(on Boundary III) and P State($0,\frac{1-\widetilde{a}}{\widetilde{a}}$).}
\label{2x-Q}
\end{center}
\end{figure}

However, there are some constraints on the physical quantities of the black hole, which can be presented in $x-\widetilde{Q}$ space. First, $x\le1$ apparently holds since the horizon radius $r_{+}$ is less than cavity radius $r_{B}$.  On the one hand, we need to ensure that the temperature $\widetilde{T}$ is not less than zero, which gives
\begin{equation}
\widetilde{Q}^2\le x^2+3\widetilde{a}\omega x^{1-3\omega}.\label{Q1}
\end{equation}
Note that $\widetilde{Q}^2=x^2+3\widetilde{a}\omega x^{1-3\omega}$ means the emergence of an extreme black hole. On the other hand, all physical quantities must be real numbers thus $f(r_B)>0$ needs to be satisfied, which gives
\begin{equation}
\widetilde{Q}^2<\frac{(1-\widetilde{a})x-x^2+\widetilde{a} x^{1-3\omega}}{1-x}.\label{Q2}
\end{equation}
The above formula should be changed into $\widetilde{Q}^2<1+3\widetilde{a}\omega$ when $x=1$. In order to simplify the calculation without loss of generality, we sets $\omega=-2/3$ and only consider the case of $\widetilde{Q}>0$ since the constraints of $\widetilde{Q}$ on $x$ are symmetric. The physically allowed regions in $x-\widetilde{Q}$ space are displayed in FIG. \ref{2x-Q}, which are divided into three categories according to the value of $\widetilde{a}$. For $\omega=-2/3$, expression for Boundary I is $\widetilde{Q}=0$, expression for Boundary II is $x=1$, expression for Boundary III is $\widetilde{Q}=\sqrt{x-\widetilde{a}x-\widetilde{a}x^2}$ and expression for Boundary IV is $\widetilde{Q}= \sqrt{x^2-2\widetilde{a}x^3}$. There is no Boundary III in FIG. \ref{2x-Q1} since it is outside Boundary IV. Also, no Boundary II is in FIG. \ref{2x-Q3} since the upper limit is $x=(1-\widetilde{a})/\widetilde{a}$ instead of $x=1$.

Our strategy to study the phase structures and transitions is as follows. First of all, we use Eq. (\ref{Phiinunitsofrb}) to express $\widetilde{Q}=\widetilde{Q}(x,\Phi)$ and substitute it into Eq. (\ref{Tinunitsofrb}) and Eq. (\ref{Gibbsfreeenergy}) respectively. Therefore we have $\widetilde{T}(x,\Phi)$ and $\widetilde{F}(x,\widetilde{T},\Phi)$. Again, substituting above $\widetilde{T}(x,\Phi)$ into $\widetilde{F}(x,\widetilde{T},\Phi)$, we get $\widetilde{F}(x,\Phi)$. With $x$ as the parameter, $\widetilde{F}-\widetilde{T}$ diagrams for fixed $\Phi$ thus can be plotted. However those $\widetilde{F}-\widetilde{T}$ curves are the cases that belong to the black hole solution. We haven't consider the cases for the boundaries. The situation of each boundary is different and we need to figure out the lowest free energy on each boundary when the grand canonical ensemble parameter $\widetilde{T}$ and $\Phi$ are fixed. 

On the Boundary I, that is, $\widetilde{Q}=0$. We notice that
\begin{equation}
\frac{\partial \widetilde{F}}{\partial \widetilde{Q}}\big|_{Boundary\text{ }I}=-\Phi<0,
\end{equation}
which implies that on the boundary other than $(0, 0)$, points with lower free energy can be found in physically allowed region since the points on the right of $(0,0)$ are beyond physically allowed region. We refer to $(0, 0)$ as the H state, which might be a candidate with the globally lowest free energy.

On the Boundary II, that is, $x=1$, which implies that the cavity merges with the event horizon. We find that
\begin{equation}
	\frac{\partial \widetilde{F}}{\partial \widetilde{Q}}\big|_{Boundary\text{ }II}=-\Phi<0.
\end{equation}
This indicates on that boundary, the free energy will decrease as $\widetilde{Q}$ increases, so the point with the lowest free energy is the right endpoint which is denoted by M State. In M State, the black hole is an extremal black hole and the cavity merges with the event horizon.

On the Boundary III, that is, $\widetilde{Q}=\sqrt{x-\widetilde{a}x-\widetilde{a}x^2}$. Free energy of this boundary can be written as
\begin{equation}
\widetilde{F}=1-\frac{ x^2\widetilde{T}}{16}-\Phi  \sqrt{x-\widetilde{a} x-\widetilde{a} x^2}. \label{freeenergyBIII}
\end{equation}
If $\widetilde{T}$ and $\Phi$ are determined, the globally minimal free energy on this boundary is also determined since it's only the function of $x$. Nevertheless, what we really need is that when $\Phi$ is fixed, the globally minimal free energy of the boundary changes with temperature $\widetilde{T}$. To solve this problem, we perform the following deduction
\begin{equation}
\frac{\partial \widetilde{F}(x,\widetilde{T},\Phi)}{\partial x}=0 \Longrightarrow \widetilde{T}(x,\Phi)=\frac{4\Phi  (2 \widetilde{a} x+\widetilde{a}-1)}{x \sqrt{x-\widetilde{a} x-\widetilde{a} x^2}}.
\end{equation}
Substitute it into Eq. (\ref{freeenergyBIII}), we obtain
\begin{equation}
\widetilde{F}(x,\Phi)=1+\frac{\Phi   (2\widetilde{a}x^{2}+3\widetilde{a}x-3x)}{64 \pi  \sqrt{x-\widetilde{a} x-\widetilde{a} x^2}}.
\end{equation}
With $x$ as the parameter, $\widetilde{F}-\widetilde{T}$ curves for fixed $\Phi$ can be plotted. It's worth noting that the above equations can only describe the behavior of Stagnation points on the boundary thus the endpoints are supposed to be considered since there is no singularity for $\widetilde{F}(x,\Phi)$ on Boundary III. For the case shown in FIG. \ref{2x-Q2}, $(1-\widetilde{a})/2\widetilde{a}<x<1$. We denote $x=(1-\widetilde{a})/2\widetilde{a}$ as N State and $x=1$ as M State and the point between M State and N State may also be the candidate of globally minimal free energy, which is denoted as L State. For the case shown in FIG. \ref{2x-Q3}, $(1-\widetilde{a})/2\widetilde{a}<x<(1-\widetilde{a})/\widetilde{a}$. We denote $x=(1-\widetilde{a})/\widetilde{a}$ as P State and L State between P State and N State may has the locally minimal free energy. In N State, the black hole is an extremal black hole and the cavity merges with the outer event horizon. In L State, the cavity merges with the outer event horizon. In P State, the charge of the black hole equals to zero and the cavity merges with the outer event horizon.

On the Boundary IV, that is, $\widetilde{Q}= \sqrt{x^2-2\widetilde{a}x^3}$. The black hole is an extremal black hole on this boundary. In the previous discussion, we have actually reached such a conclusion that Boundary IV belongs to the black hole solution and that's why Eq. (\ref{Q1}) can take the equal sign. Therefore, Boundary IV doesn't need to be considered again and the globally free energy on the boundary can only be the endpoints, namely H State or M State in FIG. \ref{2x-Q1}, H State or N State in FIG. \ref{2x-Q2} and FIG. \ref{2x-Q3}.

\begin{figure}[ptb]
\begin{center}
\subfigure[{~\scriptsize $\widetilde{a}=0.2$. From left to right, the values of $\Phi$ are $1.00, 0.30, 0.20$.}]{
\includegraphics[width=0.32\textwidth]{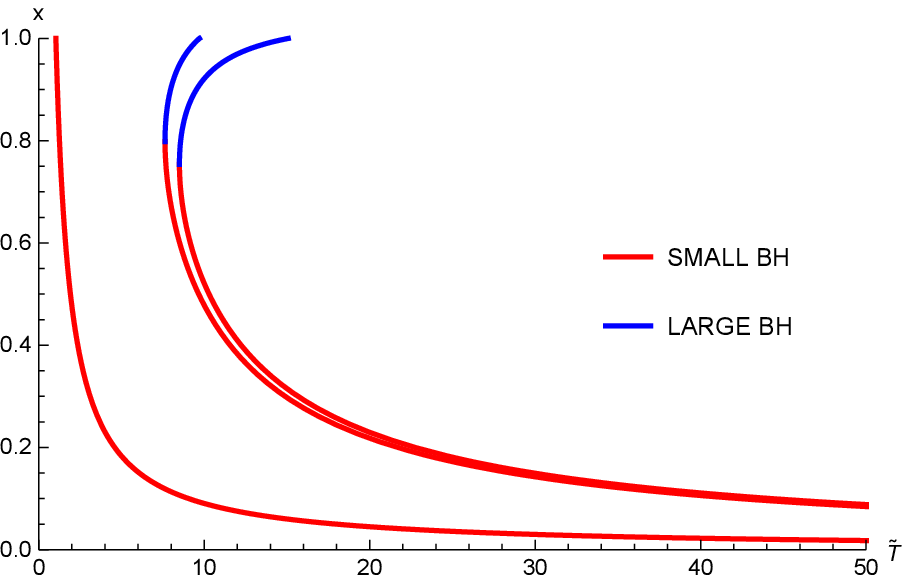}\label{2x-T1}}
\subfigure[{~\scriptsize $\widetilde{a}=0.45$. Starting from $x=0.6$, from left to right, the values of $\Phi$ are $2.00, 1.20, 1.00$.}]{
\includegraphics[width=0.32\textwidth]{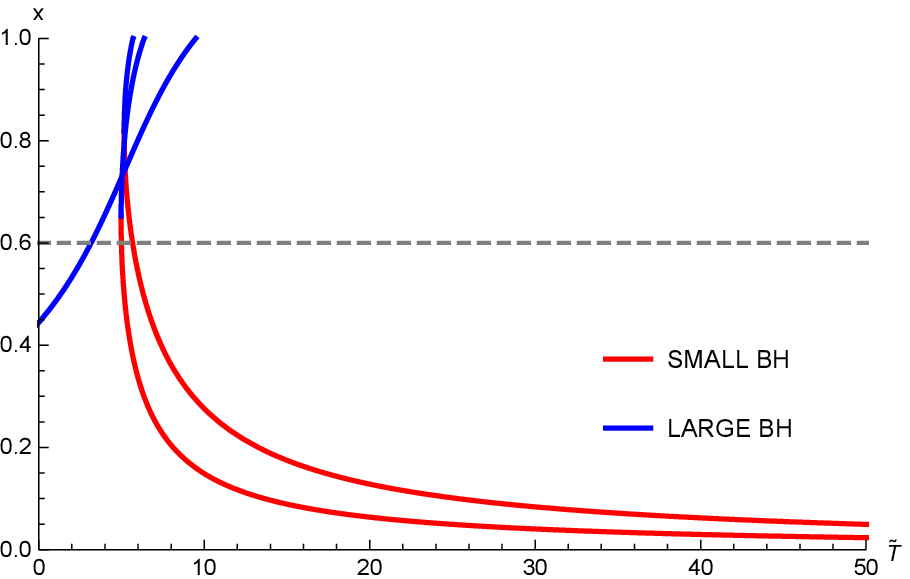}\label{2x-T2}}
\subfigure[{~\scriptsize $\widetilde{a}=0.7$. From left to right, the values of $\Phi$ are $2.00, 1.00$.}]{
\includegraphics[width=0.32\textwidth]{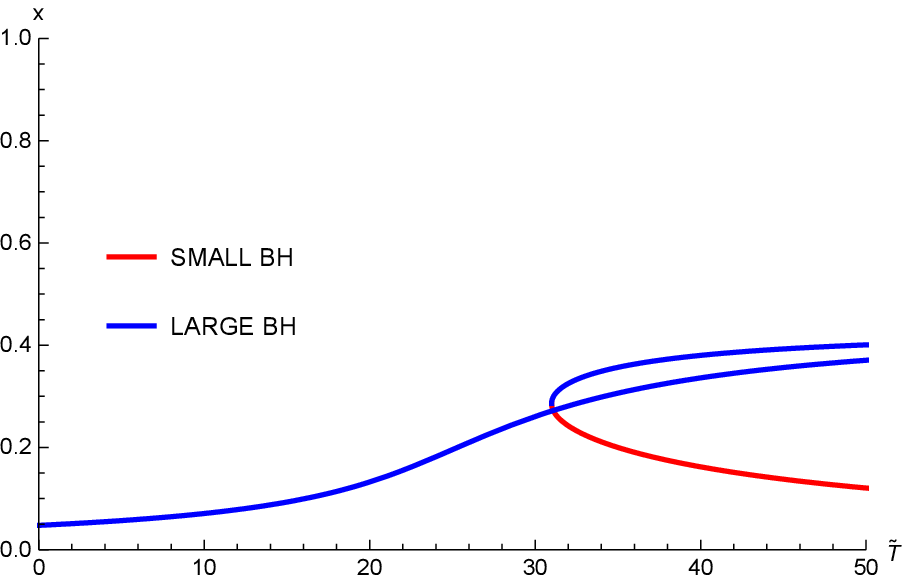}\label{2x-T3}}
\caption{~\scriptsize Plots of $x$ against $\widetilde{T}$ for different $\widetilde{a}$ and $\Phi$. The Small BH with the negative slope are unstable states while the Large BH are stable with the positive slope.}
\label{2x-T}
\end{center}
\end{figure}

Once again, we use the heat capacity at constant potential $C_{\Phi}$ to judge thermal stability. As discussed before, $C_{\Phi}>0$ and $C_{\Phi}<0$ correspond to stable and unstable states, respectively. Using $\widetilde{T}(x,\widetilde{Q})$ and $\Phi(x,\widetilde{Q})$ in Eq. (\ref{Tinunitsofrb}) and Eq. (\ref{Phiinunitsofrb}) to eliminate $\widetilde{Q}$, we obtain the expression of $x$ against $\widetilde{T}$ and $\Phi$, which is plotted in FIG. \ref{2x-T}. We divide it into three cases in advance and each case corresponds to a subgraph.

\begin{figure}
\begin{center}
\includegraphics[width=0.6\textwidth]{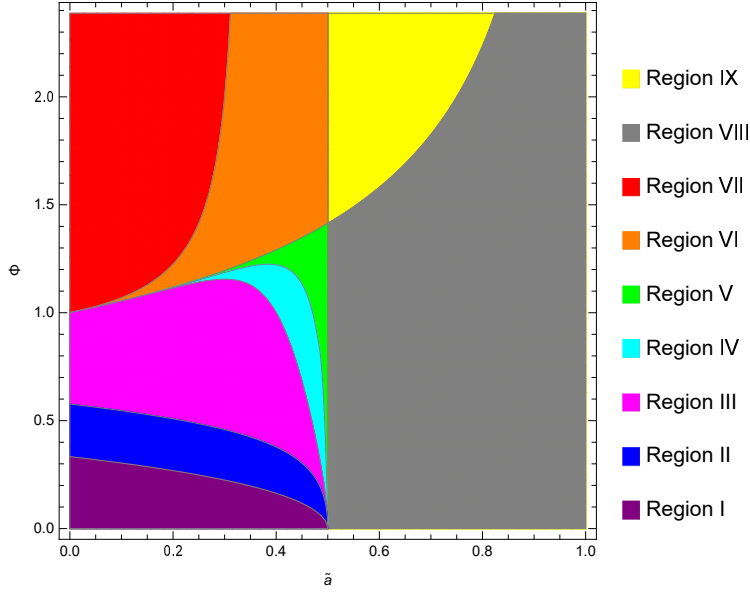}
\caption{~\scriptsize Nine regions in $\Phi-\widetilde{a}$ phase space.} \label{2phi-a}
\end{center}
\end{figure}

\begin{figure}[ptb]
\begin{center}
\subfigure[{~\scriptsize Region I: $\widetilde{a}=0.2,\Phi=0.20$.}]{
\includegraphics[width=0.32\textwidth]{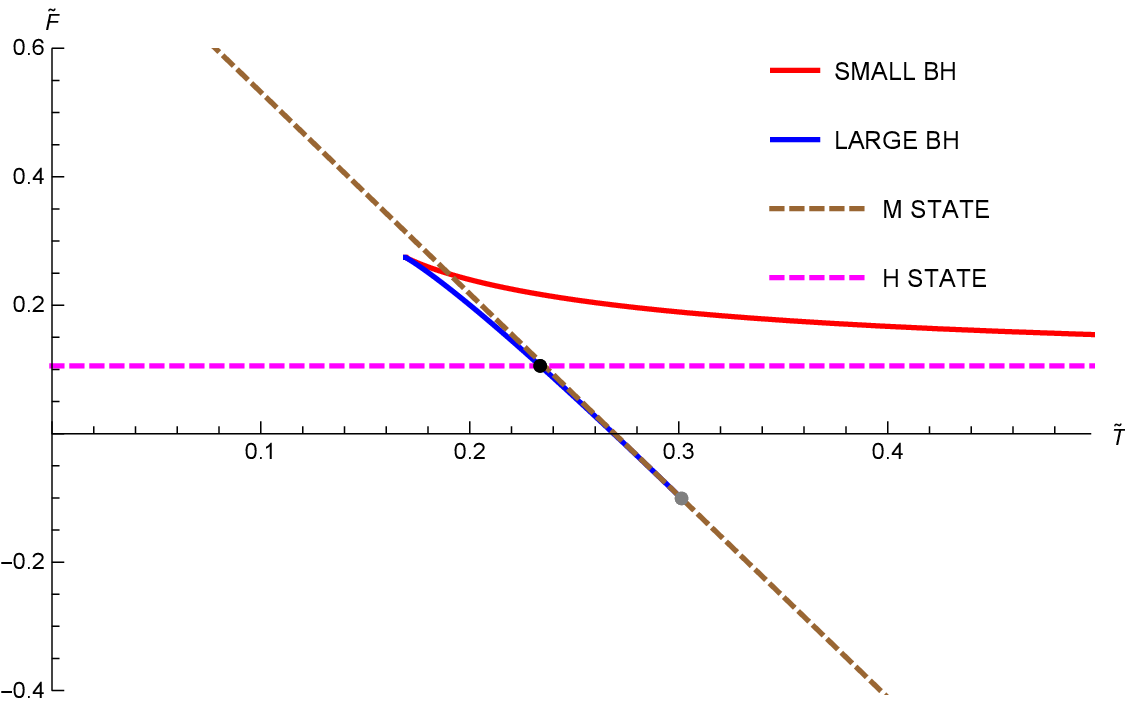}\label{2F-T1}}
\subfigure[{~\scriptsize Region II: $\widetilde{a}=0.2,\Phi=0.30$.}]{
\includegraphics[width=0.32\textwidth]{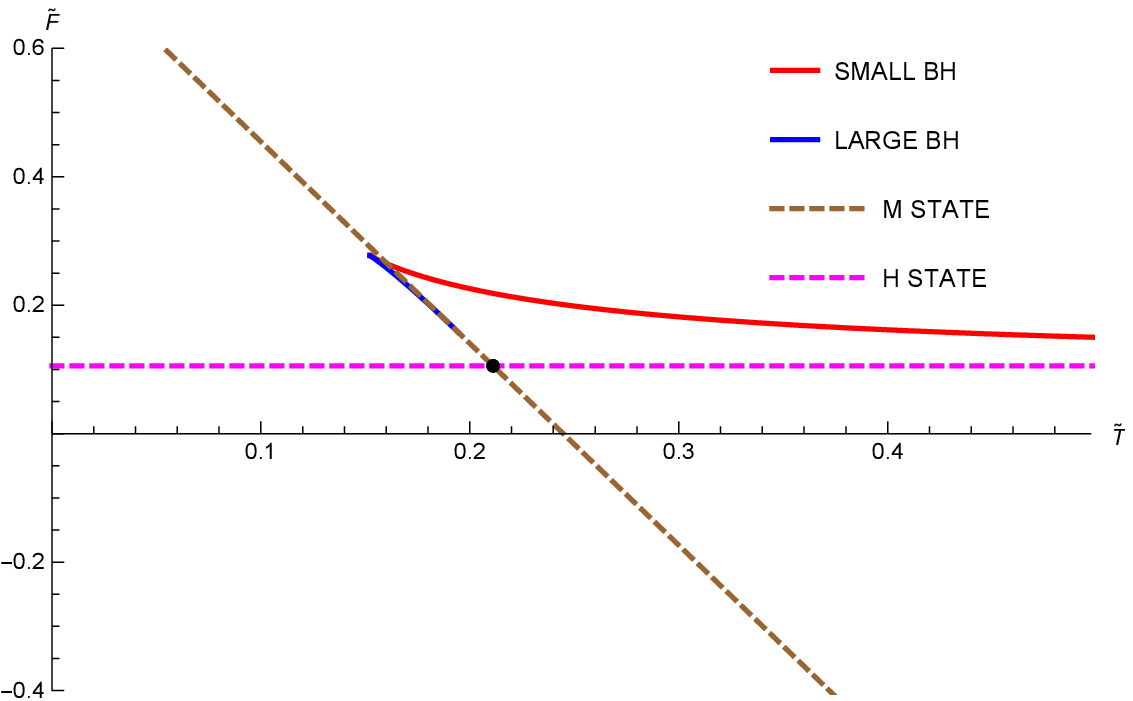}\label{2F-T2}}
\subfigure[{~\scriptsize Region III: $\widetilde{a}=0.2,\Phi=1.00$.}]{
\includegraphics[width=0.32\textwidth]{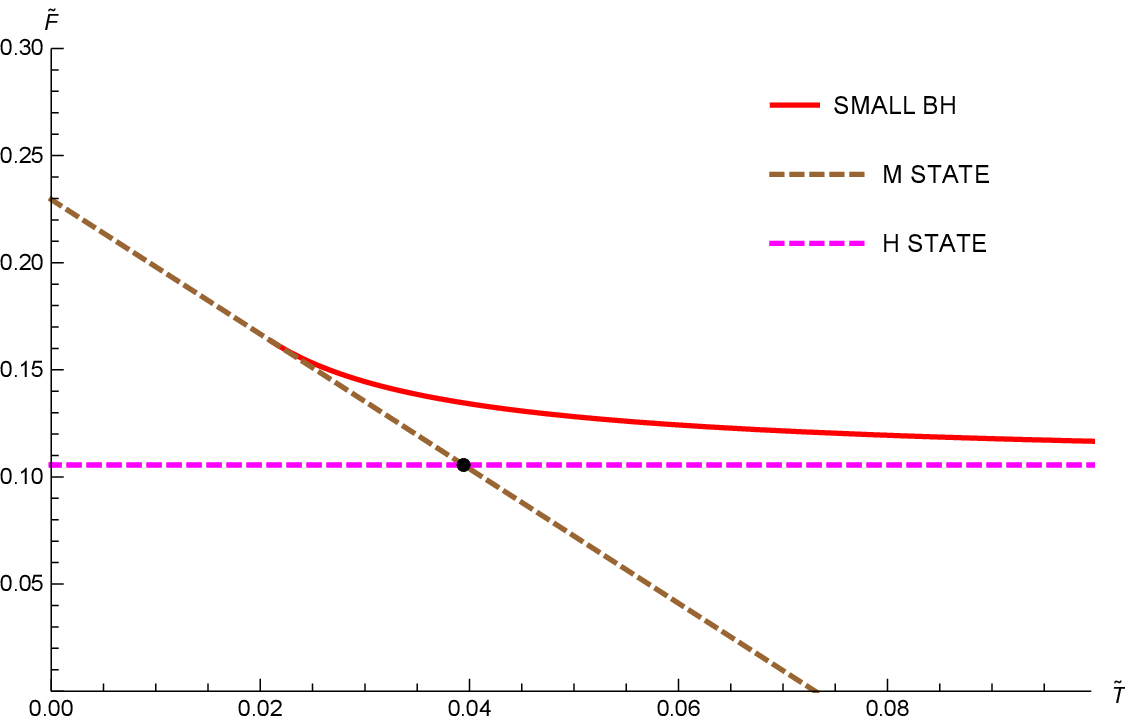}\label{2F-T3}}
\subfigure[{~\scriptsize Region IV: $\widetilde{a}=0.45,\Phi=1.00$.}]{
\includegraphics[width=0.32\textwidth]{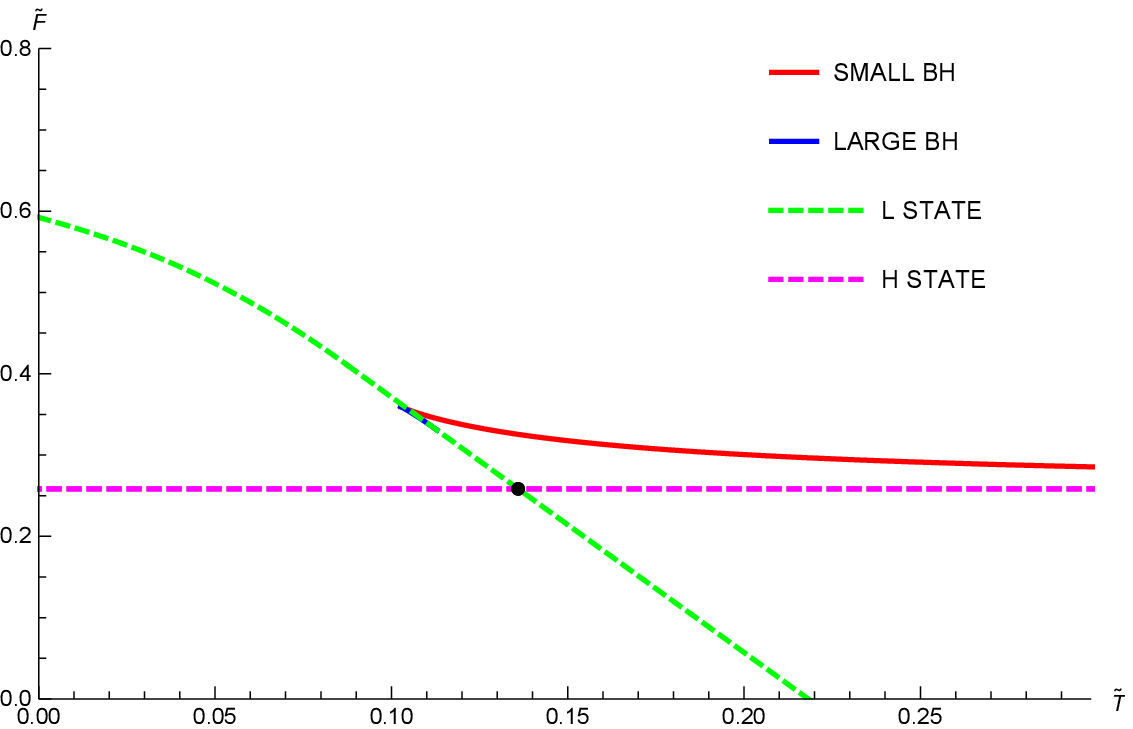}\label{2F-T4}}
\subfigure[{~\scriptsize Region V: $\widetilde{a}=0.45,\Phi=1.20$.}]{
\includegraphics[width=0.32\textwidth]{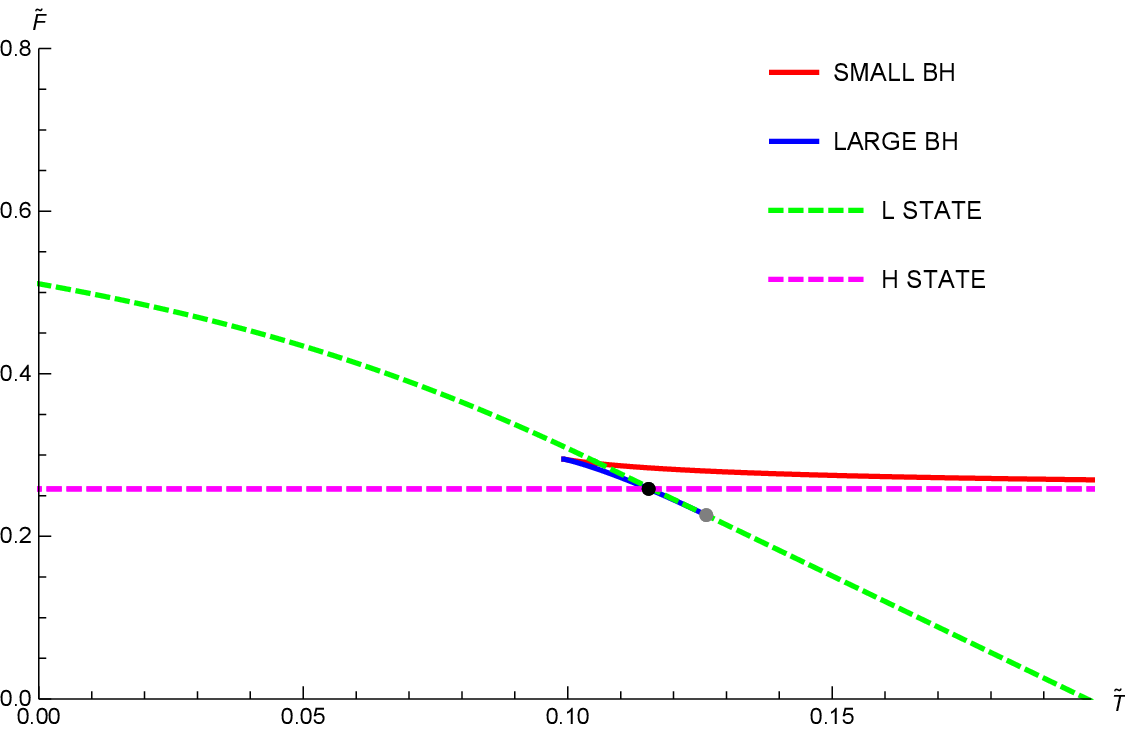}\label{2F-T5}}
\subfigure[{~\scriptsize Region VI: $\widetilde{a}=0.45,\Phi=2.00$.}]{
\includegraphics[width=0.32\textwidth]{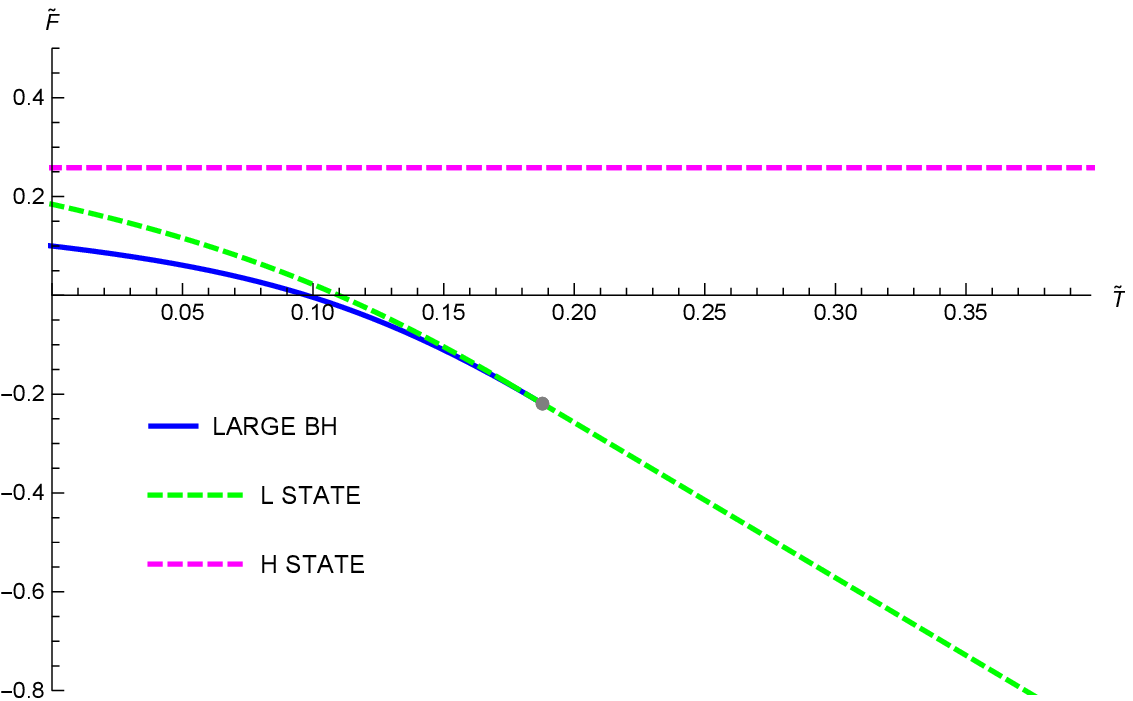}\label{2F-T6}}
\subfigure[{~\scriptsize Region VII: $\widetilde{a}=0.2,\Phi=2.00$.}]{
\includegraphics[width=0.32\textwidth]{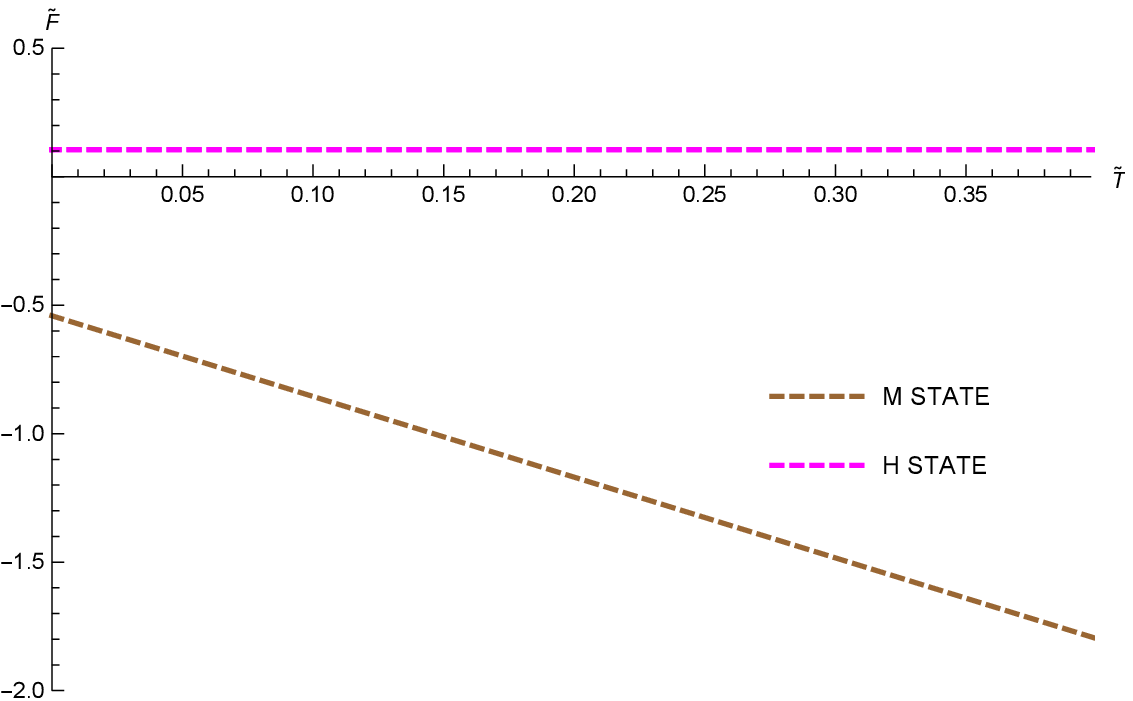}\label{2F-T7}}
\subfigure[{~\scriptsize Region VIII: $\widetilde{a}=0.7,\Phi=1.00$.}]{
\includegraphics[width=0.32\textwidth]{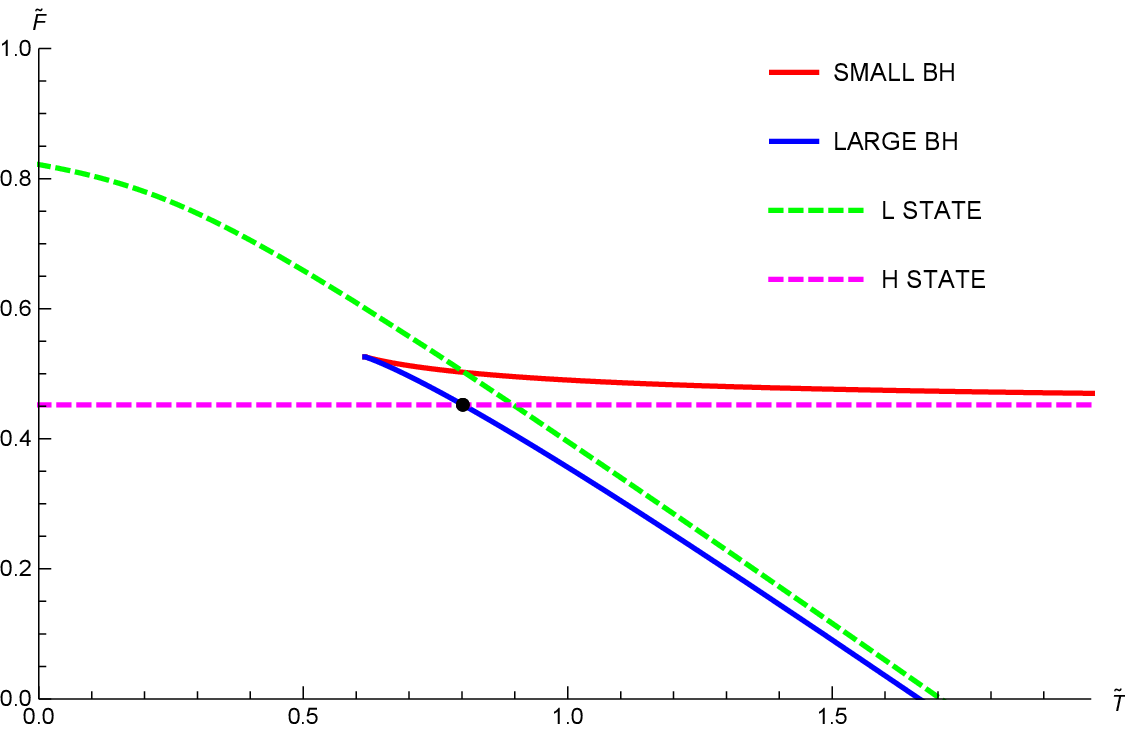}\label{2F-T8}}
\subfigure[{~\scriptsize Region IX: $\widetilde{a}=0.7,\Phi=2.00$.}]{
\includegraphics[width=0.32\textwidth]{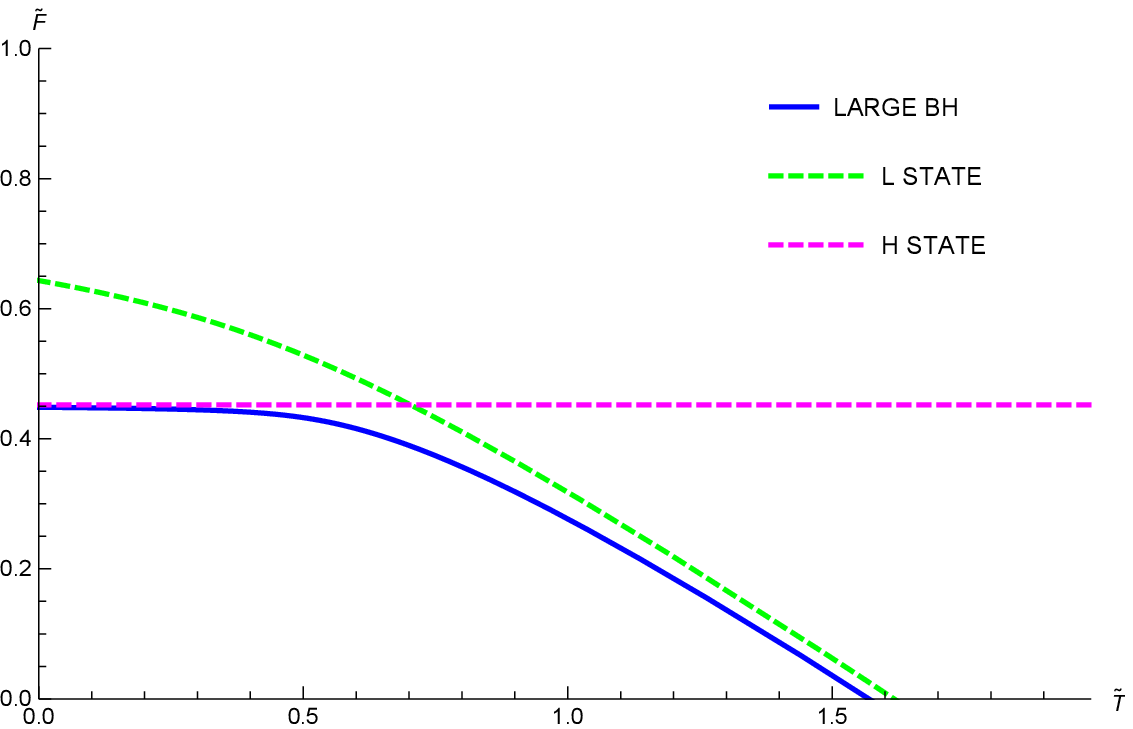}\label{2F-T9}}
\caption{~\scriptsize Plots of free energy $\widetilde{F}$ against temperature $\widetilde{T}$ in nine regions. The black dot is where a first-order phase transition occurs while the gray dot is where a second-order phase transition occurs. The solid lines represent the physically allowed region, and the dashed lines represent the boundaries.}
\label{2F-T}
\end{center}
\end{figure}

In addition, we find there are nine regions in $\Phi-a$ phase diagram, as shown in FIG. \ref{2phi-a}. To obtain the detailed information about the phase structures and transitions of each region, we display the $\widetilde{F}-\widetilde{T}$ diagrams of the physically allowed regions and the candidates on the boundaries in FIG. \ref{2F-T}. The specific and representative cases for each region are selected and the subsequent analysis are as follows.

For Region I, the $\widetilde{F}-\widetilde{T}$ diagram is plotted in FIG. \ref{2F-T1}, and the corresponding physically allowed region is shown in FIG. \ref{2x-Q1}. As discussed before, H State and M State are the candidates for minimal free energy on the boundary. Therefore, one can plot the free energy of H State, M State and the locally stationary points in the physically allowed region against temperature. There are two branches for the physically allowed region, Small BH and Large BH, which connotes that there are both a locally maximal and a locally minimal free energy in that region. When temperature $\widetilde{T}$ is small, H State has the minimal free energy thus it's the globally stable state. When temperature continues to rise, a first-order phase transition occurs from H State to Large BH and the latter becomes the thermodynamical preferred state. As $\widetilde{T}$ increases further, a second-order phase transition occurs then M State becomes the stablest state.

For Region II, the $\widetilde{F}-\widetilde{T}$ diagram is plotted in FIG. \ref{2F-T2}. Compared with Region I, only the value of $\Phi$ changes in this region, thus the candidates remain unchanged since the physically allowed region stays the same. However the difference is that the free energy of the point where Large BH tangent to M State is higher than the first-order transition, resulting in only the first-order transition from H State to M State occurs in this region.

For Region III, the $\widetilde{F}-\widetilde{T}$ diagram is plotted in FIG. \ref{2F-T3}, from which one can realize that there is only the Small BH branch in the physically allowed region. And this just corresponds to the case $\Phi=50$ in FIG. \ref{2x-T1}. As $\widetilde{T}$ increases, there is only a first-order phase transition occurs from H State to M State.

For Region IV, the $\widetilde{F}-\widetilde{T}$ diagram is plotted in FIG.\ref{2F-T4}, of which the physically allowed region is displayed in FIG. \ref{2x-Q2} and the $x-\widetilde{T}$ diagram is plotted in FIG. \ref{2x-T2}. It's distinct from the case shown in Region II mainly because the locally minimal free energy on the boundary lies at L State that is between M State and N State. We can see from FIG. \ref{2F-T4} that L State is actually a combination of a curve and a straight line. The curve corresponds to the movement of L State on Boundary III, while the straight line corresponds to L State coincides with M State. When the temperature $\widetilde{T}$ rises, the black hole undergoes a first-order phase transition from H State to L State.

For Region V, the $\widetilde{F}-\widetilde{T}$ diagram is plotted in FIG. \ref{2F-T5}. Unlike the Region IV, in this region, the free energy of the point where Large BH tangent to L State is lower than the first-order transition, resulting in the addition of a second-order transition from Large BH to L State. In general, a first-order phase transition and a second-order phase transition occur sequentially in this region.

For Region VI, the $\widetilde{F}-\widetilde{T}$ diagram is plotted in FIG. \ref{2F-T6}. There is only the Large BH branch for physically allowed region, which can also be realized from FIG. \ref{2x-T2}. As $\widetilde{T}$ increases, a second-order phase transition occurs from Large BH to L State.

For Region VII, the $\widetilde{F}-\widetilde{T}$ diagram is plotted in FIG. \ref{2F-T7} and the physically allowed is shown in FIG. \ref{2x-Q1}. What's special is that no phase transition occurs in this region. In this case, only the point that has the minimal free energy on the boundary might be the globally stable state, on account of Eq. (\ref{Tinunitsofrb}) and Eq. (\ref{Phiinunitsofrb}) without real roots in the physically allowed region. Also, it implies that it's impossible to find a $x-\widetilde{T}$ curve in FIG. \ref{2x-T1}, and that is why there are only three corresponding curves for four regions.

For Region VIII, the $\widetilde{F}-\widetilde{T}$ diagram is plotted in FIG. \ref{2F-T8}. As $\widetilde{T}$ increases, a first-order transition occurs from H State to Large BH. This region is very similar to the Region V, but the difference is that no second-order phase transition occurs since Large BH line and L State line will never intersect.

For Region IX, the $\widetilde{F}-\widetilde{T}$ diagram is plotted in FIG. \ref{2F-T9}. This region is a deformation of the Region VI, and it is very similar to Region VIII in that when $\widetilde{T}$ tends to infinity, the Large BH branch and the L State line approach infinitely but never intersect, thus no phase transition occurs.

In summary, as $\widetilde{T}$ increases, a first-order phase transition occurs in Region II, Region III, Region IV and Region VIII, a second-order phase transition occurs in Region VI, a first-order phase transition and a second-order phase transition occur successively in Region I and Region V, and there is no phase transition occurs in Region VII and Region IX. In FIG. \ref{2phi-a}, one can refer to $L_{i,i+1}$ as the critical line between Region $i$ and Region $i+1$. As $\widetilde{a}$ increases, $L_{34}$, $L_{45}$, $L_{56}$ and $L_{67}$ originate from the same $\Phi$ and gradually diverge. Meanwhile, the value of $\Phi$ on $L_{12}$, $L_{23}$, $L_{34}$ and $L_{45}$ decreases and converges to zero as $\widetilde{a}\rightarrow 0.5$. It is worth noting that the straight line $\widetilde{a}=0.5$ divides the original $L_{56}$ into the current $L_{56}$ and $L_{89}$. That is because when $\widetilde{a}>0.5$, the Large BH branches tend to infinity and never intersect with L State lines as $\widetilde{T}$ rises, as explained before. However, for the general case, if $1+3\widetilde{a}\omega<0$, M State disappears and the Large BH branches of the physically allowed region never intersect with L State lines as $\widetilde{T}$ rises, which leads to the disappearance of the second-order phase transition. Equivalently, the equation of the vertical line in $\Phi-\widetilde{a}$ diagram is
\begin{equation}
\widetilde{a}=-1/3\omega.
\end{equation}

\begin{figure}[ptb]
\begin{center}
\subfigure[{~\scriptsize $\widetilde{a}=0.45$.}]{
\includegraphics[width=0.4\textwidth]{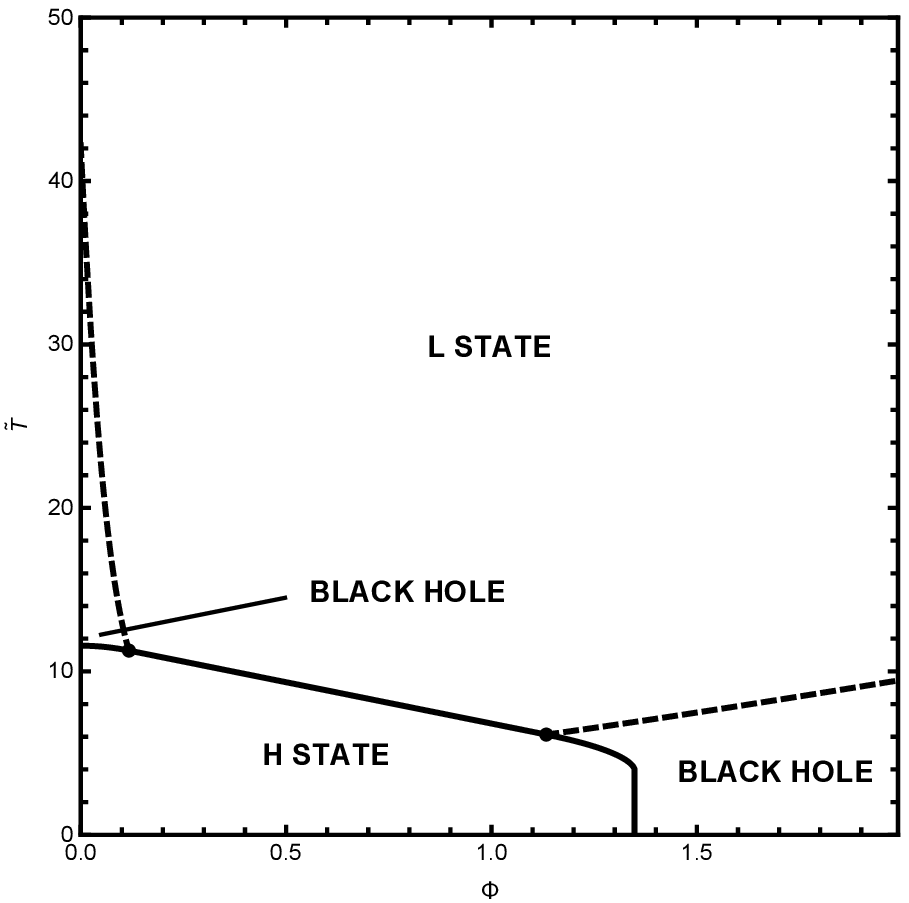}\label{2T-phi1}}
\subfigure[{~\scriptsize $\widetilde{a}=0.7$.}]{
\includegraphics[width=0.4\textwidth]{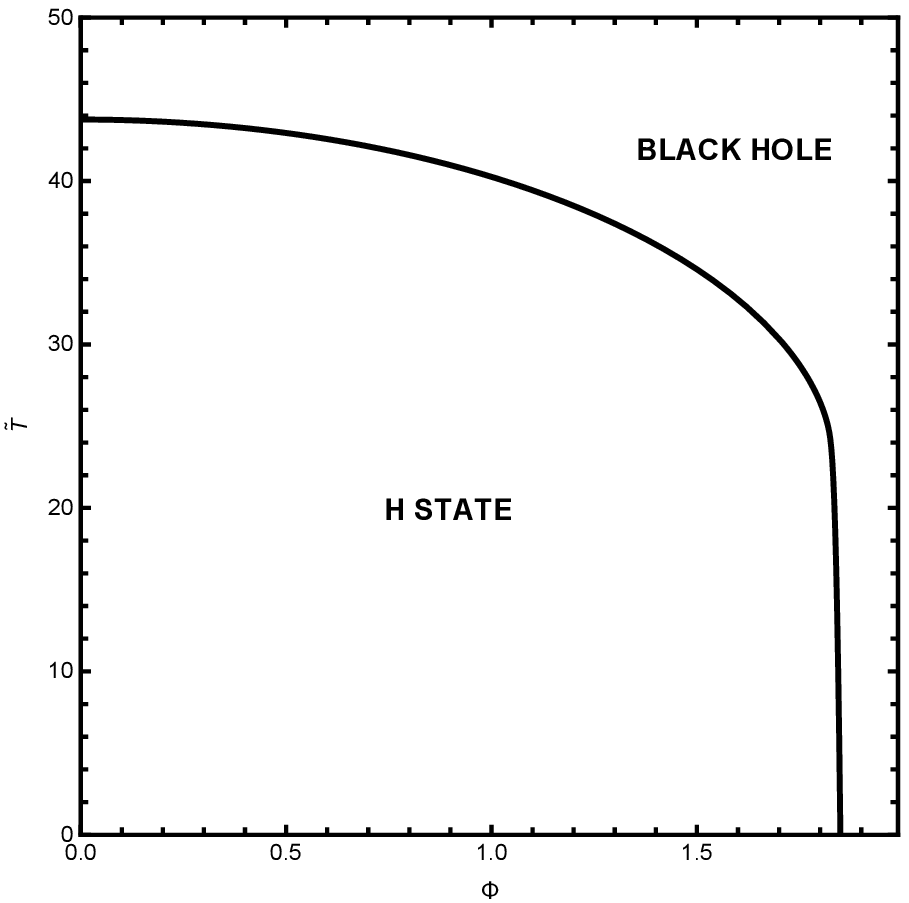}\label{2T-phi2}}
\caption{~\scriptsize The phase diagrams in $\widetilde{T}-\Phi$ space for different $\widetilde{a}$. A first-order phase transition occurs on the solid curve and a second-order phase transition occurs on the dashed curve. The black dot is the intersection of the two curves.}
\label{2T-phi}
\end{center}
\end{figure}

It's intriguing to plot the $\widetilde{T}-\Phi$ phase diagrams to conclude the conclusions above. Just as shown in FIG. \ref{2T-phi}, we set $\widetilde{a}=0.45$ and $0.7$ respectively. In FIG. \ref{2T-phi1}, there are three potential phases with the lowest free energy. When $\Phi<\Phi_{c1}\approx0.12$, H State, Black Hole and L State in turn become the globally stable state as $\widetilde{T}$ increases. When $\Phi_{c1}<\Phi<\Phi_{c2}\approx1.13$, only H State and L State have the globally lowest free energy with the change of $\widetilde{T}$. When $\Phi_{c2}<\Phi<\Phi_{c3}\approx1.35$, the three phases become the candidates of lowest free energy again. Significantly, as $\Phi$ rises to $\Phi_{c3}$, the first-order phase transition line between H State and Black Hole will become almost a vertical line. It corresponds to the transition between Region V and Region VI, that is, when $\Phi$ increases by a small amount, $\widetilde{T}$ decreases sharply. In FIG. \ref{2T-phi2}, there are only two potential phases with the lowest free energy. As $\Phi$ rises to $\Phi_{c4}\approx1.83$, $\widetilde{T}$ decreases rapidly, which corresponds to the transition between Region VIII and Region IX.

\section{Conclusion}

In this paper, we considered a grand canonical ensemble and studied the phase structures and transitions of a RN black hole surrounded by quintessence dark energy in AdS space, and in a cavity, separately. For the quintessence RN-AdS black, we fixed the temperature and the potential then found that for some temperature, there are two branches of the black hole, namely Small BH and Large BH. By calculating the heat capacity at constant potential, one discovered that the Small BH branch is unstable while the Large BH is stable. Furthermore, a first-order transition occured as temperature increases for some small values of $\Phi$ instead of large $\Phi$. Thus we divided all the cases into two regions in $\Phi-\widetilde{a}$ phase diagram in FIG. \ref{1phi-a}, in which one point can decide whether a phase transition occurs or not. Also, a $\widetilde{T}-\Phi$ phase diagram was displayed in FIG. \ref{1T-phi}.

On the other hand, we investigated the phase structures and transitions of the black hole in another boundary condition, i.e., Dirichlet wall. To begin with, we used the energy of the system obtained in \cite{Brown:1994gs} to construct the first law of thermodynamics. The case for the barotropic index $\omega=-2/3$ was studied in detail. Besides, we got the constraints of black hole parameters $x$ and $\widetilde{Q}$ in a cavity, which were separated into three cases depends on the value of $\widetilde{a}$. In the physically allowed region, we obtained the $\widetilde{F}-\widetilde{T}$ curves for fixed $\Phi$ with $x$ as the parameter. On the boundary, the position of the latent lowest free energy was evaluated through the expression of $\widetilde{F}$ and one labelled these points in $x-\widetilde{Q}$ space as different states. At last, we plotted the $\widetilde{F}-\widetilde{T}$ diagrams and got the phase transitions. To sum up, there are nine regions with different phase structures. In order to distinguish them more clearly, we plotted the nine regions in the $\Phi-\widetilde{a}$ phase diagram in FIG. \ref{2phi-a}. The $\widetilde{T}-\Phi$ phase diagrams were also plotted in FIG. \ref{2T-phi}, in which the phase structures for fixed $\omega$ and $\widetilde{a}$ could be displayed expressly. The discrepancy in the phase structures and transitions of quintessence RN black hole between the two different boundary conditions is distinct.

\begin{acknowledgments}
We are grateful to Hanwen Feng, Wei Hong, Peng Wang and Mingtao Yang for useful discussions and valuable comments. This work is supported by NSFC (Grant No.11947408).
\end{acknowledgments}

\end{document}